\documentclass[sigconf]{acmart}
\renewcommand\footnotetextcopyrightpermission[1]{}

\acmConference[MODEVAR'24]{6th International Workshop on Languages for Modelling Variability (MODEVAR 2024)}{February 6, 2024}{Bern, Switzerland}

\AtBeginDocument{%
  \providecommand\BibTeX{{%
    \normalfont B\kern-0.5em{\scshape i\kern-0.25em b}\kern-0.8em\TeX}}}

\setcopyright{acmcopyright}
\copyrightyear{2023}
\acmYear{2023}
\acmDOI{XXXXXXX.XXXXXXX}

\acmPrice{15.00}
\acmISBN{978-1-4503-XXXX-X/18/06}

\makeatletter
\newif\ifanon
\@ifclasswith{acmart}{anonymous}{\anontrue}{\anonfalse}
\makeatother

\newif\ifboth
\bothfalse

\newcommand{\ddueruemWeb}{\textsf{variability.dev}}
\newcommand{\vdev}{\textsf{variability.dev}}

\usepackage{siunitx}
\usepackage{svg}
\usepackage{tikz}
\usepackage{forest}
\usepackage{pgfplots}
\usetikzlibrary{decorations.markings, calc, positioning, backgrounds}
\tikzstyle{disabled} = [draw = gray, text = gray]
\tikzstyle{bddNode} = [circle, minimum size = 24pt, draw = black, fill = white, text depth = 0]
\tikzstyle{bddNodeSmall} = [anchor = center, circle, minimum size =16pt, draw = black, fill = white, inner sep = 2pt, text depth = 0, font = \small]
\tikzstyle{bddTerminal} = [draw = black, fill = white, inner sep = 4pt, text depth = 0]
\tikzstyle{vertex} = [circle, line width = 1pt, draw = darkgray, fill = gray, inner sep = 0, minimum size = 4pt]

\tikzstyle{fm-topdown} = [font=\footnotesize\sffamily, parent anchor = south, child anchor = north]

\tikzstyle{feature} = [rectangle, draw = black, fill = white, text = black, outer sep = 0pt, text depth = 0, text height = 1.25ex, inner sep = 1ex]

\tikzstyle{plain} = [draw=black]
\tikzstyle{mandatory} = [plain, decoration={
	markings,
	mark={at position 1 with {
			\node[circle, minimum size=4pt, inner sep = 0pt, draw=black, fill=black]{};}
	}
}, postaction = decorate]

\tikzstyle{optional} = [plain, decoration={
	markings,
	mark={at position 1 with {
			\node[circle, minimum size=4pt, inner sep = 0pt, draw = black, fill=white] {};}
	}
}, postaction = decorate]

\tikzstyle{optionalGreen} = [plain, decoration={
	markings,
	mark={at position 1 with {
			\node[circle, minimum size=4pt, inner sep = 0pt, draw = green, fill=white] {};}
	}
}, postaction = decorate]

\usepackage{lipsum}

\begin{document}

\title{variability.dev: Towards an Online Toolbox for Feature Modeling}


\author{Tobias He\ss}
\affiliation{%
  \institution{University of Ulm}
  \country{Germany}}
\email{tobias.hess@uni-ulm.de}

\author{Lukas Ostheimer}
\affiliation{%
	\institution{University of Ulm}
	\country{Germany}
}
\author{Tobias Betz}
\affiliation{%
	\institution{University of Ulm}
	\country{Germany}
}
\author{Simon Karrer}
\affiliation{%
	\institution{University of Ulm}
	\country{Germany}
}
\author{Tim Jannik Schmidt}
\affiliation{%
	\institution{University of Ulm}
	\country{Germany}
}
\author{Pierre Coquet}
\affiliation{%
	\institution{University of Ulm}
	\country{Germany}
}
\author{Sean Semmler}
\affiliation{%
	\institution{University of Ulm}
  \country{Germany}
}

\author{Thomas Th\"{u}m}
\affiliation{%
	\institution{University of Ulm}
	\country{Germany}}

\renewcommand{\shortauthors}{He\ss, Ostheimer, Betz, Karrer, Schmidt, Coquet, Semmler, and Thüm}


\begin{abstract}
	The emergence of feature models as the default to model the variability in configurable systems fosters a rich diversity in applications, application domains, and perspectives. Independent of their domain, modelers require to open, view, edit, transform, save, and configure models as well as to collaborate with others. However, at the time of writing, the top five results when googling ``Online Editor Feature Model'' point to editors that either have minimal functionality, are unmaintained or defunct, or require an offline installation, such as FeatureIDE. In this work we present a preview of our in-development online toolbox for feature modeling, variability.dev. In particular, we showcase our collaborative feature-model editor and our online configurator both of which are built on top of the FeatureIDE library.     
\end{abstract}


\begin{teaserfigure}
	\vspace*{5pt}
	\setlength{\fboxsep}{0pt}
	\centering
	\fbox{\includegraphics[width=\linewidth]{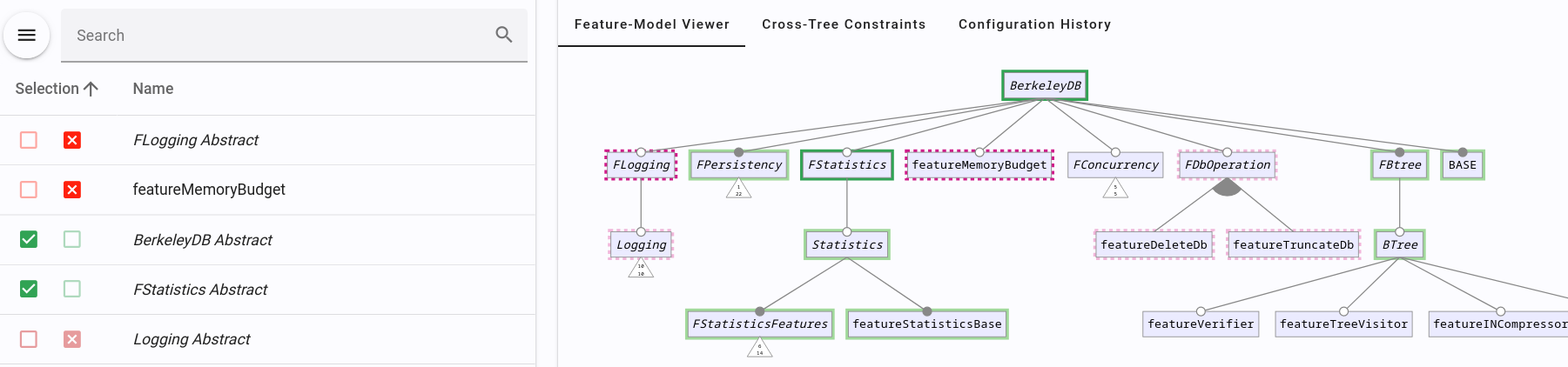}}
	\caption{Interactive Configurator of \vdev}
	\label{fig:teaserFront}
	\vspace*{10pt}
\end{teaserfigure}

\maketitle
\begin{figure*}
	
	\begin{tikzpicture}[font = \sffamily]
		\node[] (A) {\includegraphics[height = 32pt]{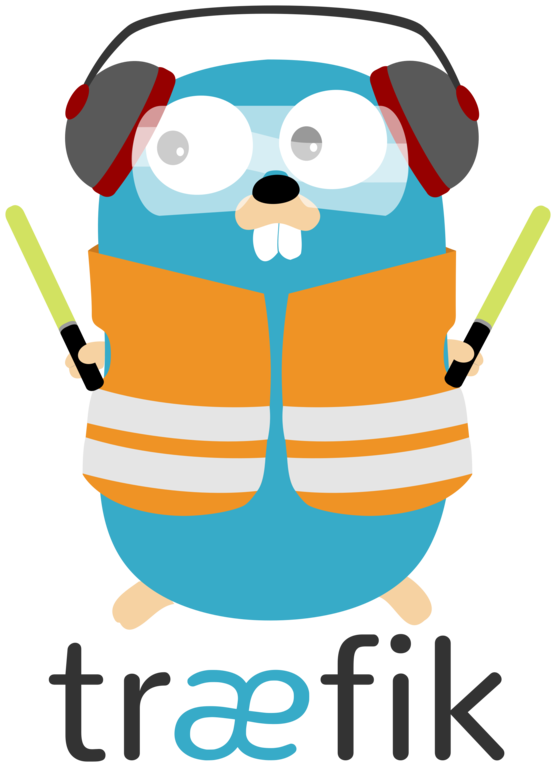}};
		\node[above = 4pt of A, text width = 2.4cm] {Reverse Proxy\\\& Load Balancer\footnotemark};

		\node[right = 2cm of A, yshift = -6pt] (B3) {\includegraphics[height = 24pt]{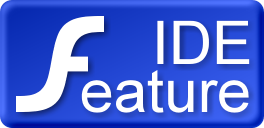}};
		\node[right = 2cm of A] (B2) {\includegraphics[height = 24pt]{img/featureIDE}};
		\node[right = 2cm of A, yshift = 6pt] (B) {\includegraphics[height = 24pt]{img/featureIDE}};		
		\node[above = 4pt of B] {FeatureIDE Service(s)};
		\node[below = 0pt of B3] {\dots};
		
		\node[right = 0pt of A.east] (H) {};
		\draw [-latex] (H.center) |- (B.west);
		\draw [-latex] (H.center) |- (B3.west);
		\draw [-latex] (H.center) |- (B2.west);
		
		\node[dashed, draw = black, rounded corners=4pt, left = 2cm of A, minimum height=3cm, minimum width=5cm] (vdev) {};
		
		\node[fill = white, xshift = -2.25cm, anchor = west] at (vdev.north) {$\vcenter{\hbox{\includegraphics[height = 16pt]{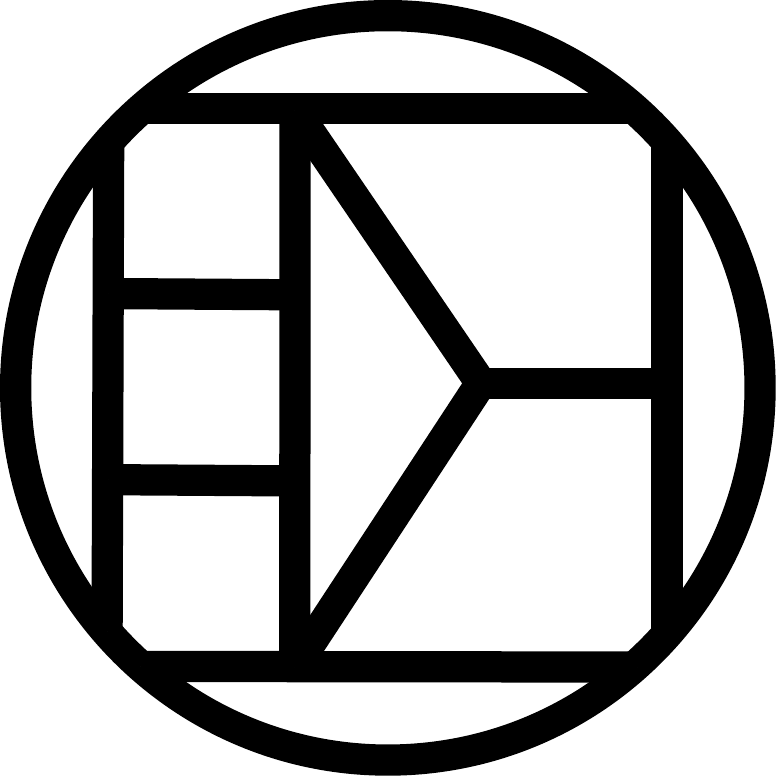}}}$ variability.dev};
		
		\node[fill = white, xshift = -2cm, anchor = west] (editor) at ([yshift = -1cm]vdev.north) {$\vcenter{\hbox{\includegraphics[height = 16pt]{img/vdev_svg-tex}}}$ Editor};
		
		\node[fill = white, xshift = -2cm, anchor = west] (configurator) at ([yshift = -2cm]vdev.north) {$\vcenter{\hbox{\includegraphics[height = 16pt]{img/vdev_svg-tex}}}$ Configurator};
		
		\node[fill = white, xshift = -1.5cm, anchor = west] at ([yshift = -2.6cm]vdev.north) {\dots};
		
		\node[left = 0pt of A.west] (H2) {};
		\draw [latex-] (editor) -| (H2.center);
		\draw [latex-] (configurator) -| (H2.center);
		
	\end{tikzpicture}
	
	\caption{Current Architecture}
	\label{fig:arch}
\end{figure*}
\section{Introduction}
\label{sec:Introduction}
Alongside the emergence of feature models as the default for modeling the variability in configurable systems, many tools and frameworks have been developed in the past two decades~\cite{MTS+17,NES:VAMOS17,KKK+:EMSE21,ACLF:SCP13,BSTRC:VaMoS07,MBC:OOPSLA09,GHF+:SPLC23,HMS+:SPLC22,RGHB:SPLC21}. Their scopes and purposes range from fully fledged IDEs such as FeatureIDEs~\cite{MTS+17} that support the entire development process of software product lines, over frameworks for feature-model analysis~\cite{ACLF:SCP13,BSTRC:VaMoS07,GHF+:SPLC23,NES:VAMOS17,HMS+:SPLC22}, to prototypes~\cite{KKK+:EMSE21} and online repositories~\cite{MBC:OOPSLA09, RGHB:SPLC21}.

Of these, variED~\cite{KKK+:EMSE21} and S.P.L.O.T~\cite{MBC:OOPSLA09} provide online feature-modeling tools. variED is a prototype of a feature-model editor with a focus on collaboration and conflict resolution~\cite{KKK+:EMSE21}. S.P.L.O.T is a repository for feature models that also provides an online feature-model editor and a configurator~\cite{MBC:OOPSLA09}, which are however cumbersome to use~\cite{RGHB:SPLC21}. In addition, S.P.L.O.T has not been maintained for almost a decade and there exists a community effort within the MODEVAR initiative to replace it in the future~\cite{RGHB:SPLC21}.

With this work, we contribute to this effort~\cite{GHF+:SPLC23,RGHB:SPLC21,LSS+:SPLC23} by showcasing our in-development collaborative feature-model editor and configurator. Both are built on top of the FeatureIDE library for means of transforming feature-model formats, anomaly detection, well-formed edit operations, and more. We envision our toolbox as a lightweight frontend to the feature-modeling functionality in FeatureIDE. However, we are currently working on supporting additional and alternative backends for analyses not supported by FeatureIDE, such as model counting.

In \autoref{sec:editor} and \autoref{sec:configurator} we showcase and discuss our editor and configurator, respectively. We discuss tools and frameworks with similar scopes in \autoref{sec:rw} and future plans for our toolbox in \autoref{sec:fw}.

\section{Overview and Usage Scenario}
\label{sec:overview}
\footnotetext{Traefik Logo by \url{https://github.com/traefik/traefik\#credits} (CC BY 3.0 Deed)}
\vdev{} is organized as a collection of dockerized microservices that may communicate with each other via restful APIs. \autoref{fig:arch} depicts the stack deployed for this showcase. All components will also function without the others, albeit with limited functionality in some cases. While we will discuss the feature-model editor in more detail in \autoref{sec:editor}, consider the following scenario for an illustration of the interaction of the various components:

User Alice opens a feature model in the editor. In order to display anomalies in the model, the editor asynchronously calls the FeatureIDE service (a restful API wrapping parts of the FeatureIDE library) to analyze the model. Possible anomalies are displayed in the model, once the request's results are available.

Alice's colleague Bob wants to view the same model that Alice is currently viewing. Alice shares her editing session with Bob via the PeerJS\footnote{\url{https://peerjs.com/}} integration. Bob finds a flawed cross-tree constraint in the model and requests edit rights from Alice. Alice grants the edit rights to Bob, who fixes the constraint. Alice notices that a previously dead feature is no longer dead, as the editor requested a reanalysis of the model after Bob's edit. Finally, Alice export and saves the fixed model.

\section{Feature-Model Editor}
\label{sec:editor}
We intended our feature-model viewer and editor to be an lightweight online alternative to editors in offline tools such as FeatureIDE~\cite{MTS+17}, with a focus on rendering speed and extensibility. Currently, our editor is capable of rendering the largest available models in fractions of a second, has functionality to efficiently navigate large models, allows for collaboration and feature-model analysis via restful calls to other services, and supports typical editing scenarios, including feature deletion by means of feature-model slicing.

\subsection{Viewing Feature Models}

\begin{figure}
	\includegraphics[width=\linewidth]{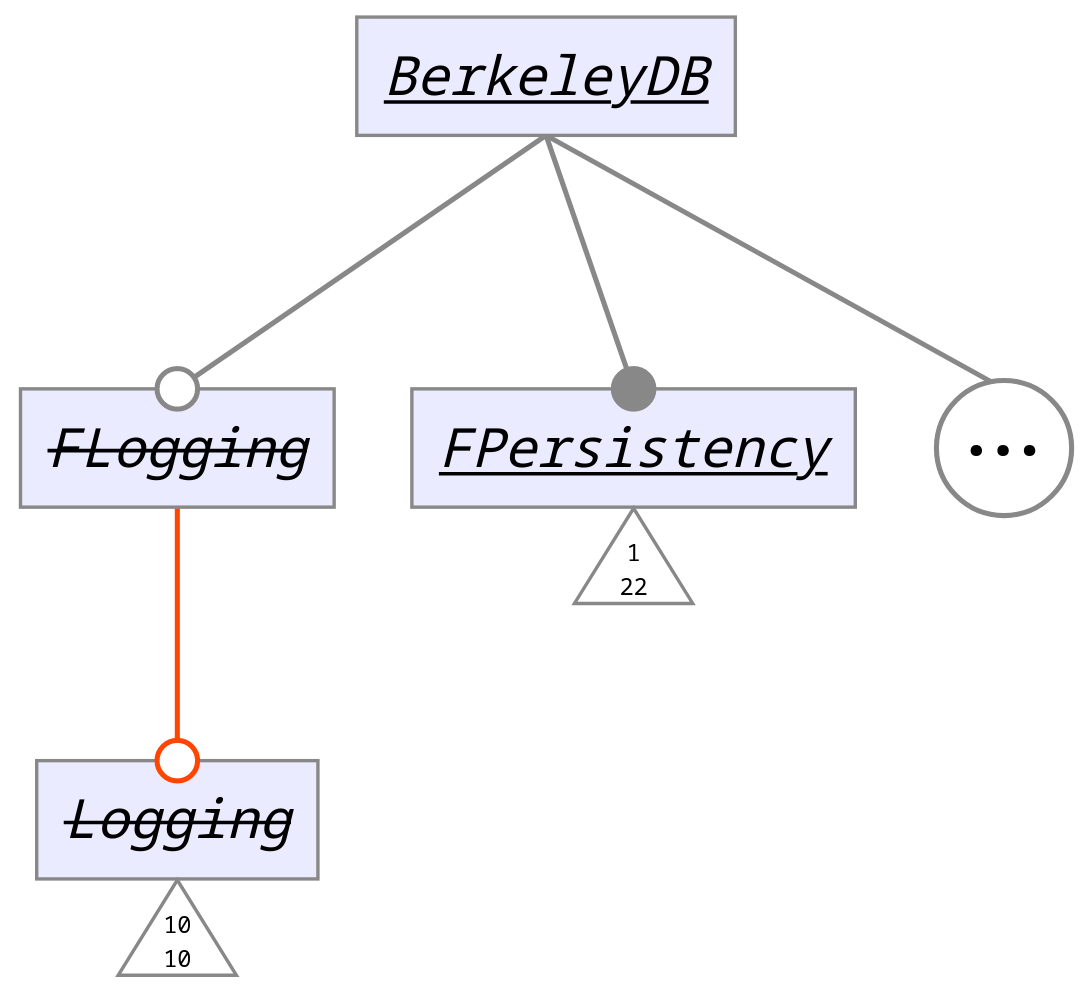}
	\caption{Horizontal and Vertical Collapsing and Anomaly Annotation in \vdev}
	\label{fig:collapsing}
\end{figure}

The feature-model viewer in \vdev{} follows the classical depiction of feature models as, for instance, in FeatureIDE~\cite{MTS+17}. Features are denoted by rectangles, with abstract features~\cite{TKES:SPLC11} typed in italic. Core features (features that are selected in every valid configuration~\cite{BSRC10}) are underlined and dead features (features that are not part of any valid configuration~\cite{BSRC10}) are stroked through. Mandatory and optional child features are depicted with the usual edges, alternative and or groups with the typical cone at the parent. Opposed to FeatureIDE, we opted to mark the edge of an false-optional feature (an optional feature that must always be selected together with its parent due to cross-tree constraints). In \autoref{fig:collapsing}, \textsf{BerkeleyDB} is an abstract and core feature, \textsf{FLogging} is abstract, optional, and dead, and \textsf{Logging} is abstract, false-optional, and also dead.

While even the largest models are rendered well below \SI{200}{\milli\second} by our viewer, large models are likely to visually overwhelm the user. Therefore, our viewer supports two types of collapsing, depicted in \autoref{fig:collapsing}. Sibling features and nodes on the same level can be hidden and are replaced by \tikz[baseline=-\the\dimexpr\fontdimen22\textfont2\relax ]{\node[circle, draw = gray, line width=1pt, inner sep=2pt] {\huge ...}; } to denote that something is hidden. Sub trees below can be hidden and are replaced by triangles as in \autoref{fig:collapsing}. The top number of the triangle refers to the hidden nodes directly below and the second number refers to the total number of hidden features below.

In order to navigate large models, we implemented a search that suggests matching features ordered by their Levenshtein distance. For visual guidance, not only the found feature is highlighted, but also the path from the root feature to the found feature. 

Cross-tree constraints are listed similarly to the constraint view in FeatureIDE. Features appearing in constraints may be selected and are highlighted with a colored border. The interaction of multiple cross-tree constraints can be illustrated by selecting multiple cross-tree constrains, as the colored borders around a feature stack.

\subsection{Editing Feature Models}
We support the typical feature model operations, namely adding, moving, and deleting features, as well as adding, editing, and deleting cross-tree constraints. Features may be added via the context menu or, with quick edits enabled, by clicking highlighted areas around features. Features may be moved laterally or arbitrarily, if semantic drag and drop is enabled. As the order of sibling features may encode semantics~\cite{FAR:SPLC20}, moving of features via drag and drop can also be disabled entirely.

For feature deletion, we employ the slicing functionality in FeatureIDE~\cite{KST+:SPLC16,ACLF:ASE11}. This ensures that cross-tree constraints remain meaningful after feature appearing in the are removed from the model and handles non-leaf features.

\subsection{Collaborative Editing}
\label{ssec:collab}
Similar to variED~\cite{KKK+:EMSE21}, our editor allows for multiple users to collaboratively edit, analyze, and view the same feature model. To circumvent the challenges incurred by simultaneous edit operations, we, opposed to variED~\cite{KKK+:EMSE21}, opted to only allow one user to edit the model at a time. This allows us to use the full set of edit operations known from FeatureIDE, including the use of feature-model slicing when deleting features.

An editing user may share his editing session via link or QR code. After joining, users can control their view of the model independently of the host's and ask to claim edit rights at any time. Both host and the currently editing user can grant edit rights, with the host being additionally able to reclaim edit rights at any time. All users share the edit history and are therefore capable of undoing or redoing edit operations.

\subsection{Import and Export}

Our feature-model editor natively supports the FeatureIDE XML format, other feature-model formats are translated using the FeatureIDE service. Currently, edited models can be exported to the formats available in FeatureIDE, including DIMACS and UVL. As our editor is based on SVG, the model can also be exported as SVG.

\subsection{FeatureIDE Integration}

To harness the functionality of FeatureIDE, we built the FeatureIDE service\footnote{\url{https://github.com/OBDDimal/FeatureIDE-Service}}, a command-line interface and restful API around the FeatureIDE library. This service currently supports the transformation of feature-model formats, anomaly detection, decision propagation, and t-wise sampling with YASA~\cite{KTS+:VaMoS20}. Requests to the restful API are handled asynchronously and are lazily loaded into the editor, once available.

\section{Feature-Model Configurator}
\label{sec:configurator}
Besides the feature-model editor, we are also developing an online feature-model configurator, whose user interface is depicted in \autoref{fig:teaserFront}. The configurator reuses many parts of the editor, for instance the feature-model viewer and the constraint view. Features may be configured from the list of features, or by clicking on features in the feature model. Every configuration decision may be rolled back using the configuration history. Selection, deselection, or freeing of a feature invokes the built-in decision propagation which selects and deselects features that are implied by the current partial configuration.

When the FeatureIDE service is available, the configurator uses the decision propagation of FeatureIDE which also marks open features, for instance when the parent of an or group is selected but none of the child features have been selected so far. Core and dead features are selected and deselected automatically when loading the model, respectively. Users may disable decision propagation and configure freely but potentially invalid. For the future, we are planning to use the explanation and configuration reparation functionalities from FeatureIDE as well. Furthermore, we are currently testing the incorporation of model counting information into the configuration process.

Like the default feature-model view, we designed the configuration view to be as inclusive as possible with regard to colorblind people. Therefore, selected features are always depicted with a solid border and deselected features with a dashed border. Only the distinction between explicitly (dark) and implicitly (light) selected or deselected features is made by color. In the future, we plan to collapse the feature model based on configuration, for example, that sub trees below a deselected feature will always be collapsed and sub trees below a selected feature will be revealed upon selection.

\section{Related Work}
\label{sec:rw}
In this section, we present work that is related to ours, in particular, previous and ongoing efforts to provide the product-line community with a platform facilitating collaboration.

\textsf{S.P.L.O.T.}\footnote{\url{http://www.splot-research.org/}} is a web-based platform providing a feature-model repository, functionality for editing and analyzing feature models, as well as an interactive configurator~\cite{MBC:OOPSLA09}. However, \textsf{S.P.L.O.T.} has not been maintained or updated in almost a decade and the models in its repository are not representative of current real-world feature models, with the majority being insignificantly tiny~\cite{RGHB:SPLC21}. Its automated analysis functionality is rather cumbersome to use (e.g., blocking the page on computation) and does not scale to current models. Most notably, analysis results are not stored or cached. Similarly, we find that their feature-model editor and configurator are not intuitive to use for users that are familiar with, for instance, FeatureIDE.\\

\textsf{FeatureIDE}\footnote{\url{https://featureide.github.io/}} is an Eclipse-based framework for feature-oriented software development~\cite{MTS+17}. In addition to feature modeling, FeatureIDE supports the entire scope of software product-line development~\cite{MTS+17}. As part of \vdev{}, we built an restful API around the FeatureIDE library and use many of its functionality with regards to feature modeling.\\

\textsf{variED}\footnote{\url{https://github.com/ekuiter/variED}} is a prototype of a collaborative feature-model editor, were multiple users may edit the same model concurrently~\cite{KKK+:EMSE21}. Resulting conflicts are automatically detected and users are provided with conflict-resolution strategies. As discussed in \autoref{ssec:collab}, the collaboration feature in \vdev{} circumvents the issues incurred by concurrent edits, by only allowing one user in the collaborative session to edit the model. This enables edit operations, such as feature deletion by means of feature-model slicing that are not covered by variED.

\section{Future Work}
\label{sec:fw}
Besides quality-of-life improvements, both already planned and as an reaction to feedback from the community, we plan to extend the functionalities of \vdev{} as follows.

\paragraph{Analysis Integration} Currently, feature models are displayed in the common format in a view that highlights anomalies in a colorblind-friendly visualization (cf. \autoref{sec:editor}). In the future, we envision additional views on a feature model that, for instance, color the features with respect to their commonality~\cite{Sundermann20}, other model-counting metrics~\cite{SNB+:VaMoS21}, or syntactic information, such as constraint interaction. Besides such views, we are currently integrating feature-model fact labels~\cite{HGP+:SPLC22} into the editor.

\paragraph{Knowledge Compilation} In addition to downloading the edited model in the various file formats or as SVG, we want to provide functionality to also download analysis results, samples, and knowledge-compilation artifacts such as BDDs or d-DNNFs in the future. Likewise, we plan to enhance the configurator with BDD-guided configuration~\cite{BK:Informatik17} and to visualize configurations in general, not only with the model but also on graphical representations of BDDs and d-DNNFs.

\paragraph{Textual Feature-Model Editor} Our editor renders even the largest models in under \SI{200}{\milli\second} on entry-level machines. In addition, it possesses functionality to quickly navigate to features of interest and to hide parts of a model. Nevertheless, there are benefits to editing a model in a textual representation, such as UVL~\cite{SFE+:SPLC21}. Stemming from the MODEVAR initiative, UVL is designed to be read and written by humans~\cite{SFE+:SPLC21, SFG+:SPLC22} aided by a language server protocol~\cite{LSS+:SPLC23}. We aim to support textual editing of feature models in our editor and especially an integration of UVL and its language server.

\section{Conclusion}
\label{sec:conclusion}

We presented two in-development tools for feature-model editing and configuration and the underlying architecture. Our editor already supports typical editing operations and is capable of handling the largest currently available models.

We presented \ddueruemWeb{}, our proposal for a web-based, collaborative platform to exchange feature models, analysis results, and related artifacts. However, our platform is not merely a static repository but every shared feature model is converted into other formats, analyzed, and sampled, when possible. Furthermore and to the best of our knowledge, \ddueruemWeb{} is the first platform to treat the evolution history of feature models as first-class citizens, an aspect on which we are planning to expand heavily in the future. Last not least, with features like the mobile-friendly feature-model viewer and editor, with its live collaboration, we also make an effort to lower the barrier for elementary product-line education.

\section*{Acknowledgments}
The authors thank Mathieu Acher, David Benavides, Rick Rabiser, Chico Sundermann, and other members of the MODEVAR initiative~\cite{BRBA:SPLC19} as well as many participants of SPLC'21, SPLC'22, FOSD'21, and FOSD'22 for many fruitful discussions and insightful comments. In particular, we want to thank Elias Kuiter for sharing his experience with collaborative editing and feature modeling in general.

\texttt{variability.dev} was and is developed in software engineering projects at the University of Ulm. We thank Sebastian Budsa, Sean Duft, Raphael Dunkel, Ruben Dunkel, Lukas Harsch, Niklas Hoehne, Tobias M{\"u}ller, and Christopher Vogel for their diligent work in their respective projects. Special thanks go to Alexander Raschke for organizing the software-engineering projects at the University of Ulm.

\bibliographystyle{ACM-Reference-Format}
\bibliography{additions.bib,MYshort.bib,literature-cleaned.bib}

\end{document}